\documentstyle[preprint,prl,aps]{revtex}

\tighten

\begin{document}

\draft

\preprint{LA-UR-93-3594 {\large\bf /} MA/UC3M/8/93}

\title{An aggregation model for electrochemical deposition \\
and its application to CoP dendritic growth}

\author{Angel S\'anchez}

\address{Escuela Polit\'{e}cnica Superior, Universidad Carlos III de
Madrid,\\ E-28913 Legan\'{e}s, Madrid, Spain\\
and\\ Theoretical Division and Center for Nonlinear Studies, Los Alamos
National Laboratory,\\ Los Alamos, New Mexico 87545}

\author{M.\ J.\ Bernal and J.\ M.\ Riveiro}

\address{Dpto.\ de F\'\i sica Aplicada, Facultad de Ciencias
Qu\'\i micas, Universidad de Castilla-La Mancha, E-13004 Ciudad Real,
Spain}

\date{October 21, 1993}

\maketitle

\begin{abstract}
We introduce an aggregation model for general electrochemical deposition
experiments.  Its most relevant feature is that it includes the overall
effect of strong applied electric fields and therefore applies to
non-equilibrium situations.  We compare our model to experiments on CoP
dendritic growth with very good agreement: The model accurately
reproduces the dependence on the current of the alloy's composition,
morphology, and growth time.
\end{abstract}

%\vspace{1cm}

\pacs{Ms.\ number \phantom{LX0000.} PACS numbers: 68.70.+w, 82.20.Wt,
05.40+j, 61.50.Cj}

\narrowtext

In the decade elapsed since the diffusion-limited-aggregation (DLA)
model \cite{DLA} was claimed to describe experimental electrochemical
clusters \cite{Matsushita}, it has been recognized that DLA is only a
crude approximation whose validity is limited to some specific cases
(basically, quasi-stationary situations).  Indeed, electrodeposits
exhibit a large variety of shapes depending on electrolyte concentration
and applied voltage; not only fractal patterns have been reported, but
also dendritic, open, and dense ones \cite{Sawada,Grier}.  A
comprehensive theoretical approach to this problem involves many factors
\cite{Libro}: It has been recently shown that if all the relevant fields
are considered (cation and anion concentration gradients, electric
field, and convection) it is possible to obtain good agreement with
experiments \cite{Fleury}.  However, these methods are based on some
coupled partial differential equations, which implies the drawback of
lengthy and complex computations.  In view of this, it would be very
useful to have a discrete counterpart for electrochemical deposition
(ECD), i.e., an aggregation model that is also able to reproduce
experimental results.

The aim of this Letter is twofold: First, to devise the simplest
aggregation model that can explain general ECD experiments; and, second,
to apply it to account for recent experiments on growth of
cobalt-phosphorus alloys by ECD \cite{MJ}.  For this last purpose the
model will include two components, but it can be applied to one
component processes as well.  Little work has been done so far on two
component ECD. To our knowledge only Nagatani and Sagu\'es addressed
this issue \cite{Nagatani}, studying from a theoretical point of view a
two-component multiparticle \cite{Voss} DLA model.  In particular, we
are not aware of any comparison between experiments and models of
two-component aggregation previous to the one we present below.  For the
sake of clarity, we first summarize the results we try to recover from
our model (see \cite{MJ} for a more detailed report).  Afterwards, we
will proceed to define the model itself, discuss the underlying physics,
and present our results.

Co$_{88}$P$_{12}$ amorphous alloys for use as soft magnetic materials
were produced by dendritic growth on the edge of a Cu substrate by ECD.
As the aim of the research was to design an efficient way to obtain
these alloys, applicable in industry, high density currents were used
instead of the low ones usually employed in more basic research.  The
initial bath composition was 75\% Co and 25\% P ions; other slightly
different compositions led essentially to the same results.  It was
observed that when the current density $J$ was increased the cobalt
fraction in the composition of the alloys increased as well, until it
saturated at 88\% Co atoms for $J\geq 4\times 10^4$ A m$^{-2}$.  Three
regions with different types of electrolytic growth were found (see
Fig.\ \ref{Composition}).  For small ($0\leq J \leq 10^4$ A m$^{-2}$)
$J$ values, the growth was planar and the Co concentration increased
with $J$.  Intermediate $J$ values ($10^4$ A m$^{-2} \leq J \leq
4\times 10^4$ A
m$^{-2}$) gave rise to dendritic morphologies, and again to an increase
of Co concentration with $J$.  Finally, for $J\geq 4\times 10^4$ A
m$^{-2}$, the growth remained dendritic but the composition reached a
constant value independent of $J$.  In this regime, deposits were always
homogeneous, even if large current density fluctuations occurred.  This
was checked by preparing samples with different deposit times (from 10
to 60 minutes).  Another interesting remark on the alloys morphology is
that the larger $J$, the more and the greater the empty spaces in the
obtained dendritic sample.  Finally, an inverse linear relation between
the time $T_{\scriptstyle\rm M}$ needed to grow a given mass $M$
of alloy and the current used to prepare it was found, i.e.,
$T_{\scriptstyle\rm M}\sim J^{-1}$.

We now turn to the aggregation model we propose to reproduce this
phenomenology.  It is clear that the physical situation correspondent to
the above experiments is that of a two-component diffusion (Laplacian)
problem with a perturbation term due to the electric field, plus another
extra term coming from the interaction between the two species during the
deposition process ({\em induced codeposition}, see below, see also
\cite{brenner}).  Other terms can possibly have some additional
influence, mainly around the cathode \cite{Fleury}.  The algorithm we
propose to mimic this continuum problem is the following.  We start from
a number of random walkers (``ions'') randomly placed on a square
lattice, with the same composition with respect to the bath than the
experimental one ($\sim 10\%$ sites occupied), and in the same relative
compositional ratio (three times more Co than P). Bidimensional lattices
were chosen to avoid large computational times while keeping a
significant number of walkers.  The lattice boundary was rectangular
(similar to the experimental one), and we took its lower side to be the
cathode.  Boundary conditions were periodic in the direction parallel to
the cathode and reflective in the perpendicular one (with the
possibility to stick to the cathode, of course).  Simulations were
done on a $300\times 400$ lattice; different sizes did not modify the
results, so we took this as the basic one in order not to spend much
CPU time.  The initial condition
evolves in time according to the following rule.  Every time step, a
walker is chosen at random (all of them with equal probability), and one
of its four nearest neighbors is chosen to be its destination site.
This is done according to the following probability: 0.5 to move
horizontally, i.e., parallel to the cathode, $0.25+p$ to move vertically
downwards (towards the cathode), and $0.25-p$ to move vertically
upwards.  We hereafter call $p$ the Co {\em bias} for reasons that will
become clear in the next paragraph.  After choosing a site to move to,
motion takes actually place if there is no other walker at that precise
node.  In case the chosen site belongs to the growing aggregate (in the
first time step, the lower layer of the lattice), the walker's previous
position is added to the aggregate with probability $s$, otherwise the
walker stays there (and can move in future time steps) with probability
$1-s$.  We term $s$ the {\em sticking probability} (in aggregation model
jargon is often called the noise reduction parameter) and again postpone
its discussion to the next paragraph.  If the walker is added to the
aggregate, a new one is created at a random site, of the same nature as
the one that has clung to the cluster \cite{otranota}.  This completes a
time step, and the definition of the algorithm.

We now discuss the physical reasons for our choice of the algorithm.
First of all, multiparticle DLA \cite{Voss} is needed to
include the consequences of having two components simultaneously.
Second, we have to introduce the large electric current.
This we do by means of $p$, the fundamental parameter of the model.
Physically, it represents the effects of the potential difference
between the electrodes and the deviations from the diffusion equation
around the cathode (see \cite{Libro,Fleury}).  It seems reasonable that
the larger the current density in the experiment, the larger $p$ must be
in the model to reproduce it.  Moreover, $p$ allows to account for the
above mentioned induced codeposition.  It is known \cite{brenner} that P
alone would not deposit, and if it does is due to the influence of other
metallic ion present in the solution (Co in our case). The simplest way
to  take this into account is to relate the bias for Co walkers to the
bias for P walkers, in the form $p_{\scriptstyle\rm Co}= p,\>
p_{\scriptstyle\rm P}=\alpha p$; in this fashion, Co ions
will drag P ones towards the cathode with a strength
proportional to their one attraction to the negative electrode.  The
$\alpha$ parameter will be determined below.  The remaining
physico-chemical complicated processes, difficult to quantify, are
included in the parameter $s$.  In particular, notice that $s$ is
related to the activation energy of the deposition process itself, and
that it should be easier for faster ions to deposit.  If $s$ is to
represent this and other effects, it must certainly be related in some
way to the current, i.e., to $p$; moreover, $s$ should grow with $p$
\cite{nota}.  We adopted this hypothesis and we consequently did not
choose $s$ independently but, rather, established a relationship between
both parameters.
Since the exact relationship is of course unknown, we
took $s$ in the interval $0<s\leq 1$ for each value of $p$. We found
that the resulting concentration value was rather insensitive to
the precise value chosen for $s$ (see further discussion of this
parameter below).
This is very important, because otherwise
the model would not be of much interest as this relation would have to
be determined for every particular case.  Thus, to summarize, we want to
stress that the main model parameter is $p$, and all others must be
related to it to explain the corresponding physical
phenomena in a natural way.

We have performed a detailed simulation program of the above model.
For simplicity, we started by fixing $\alpha=0$, i.e., Co ions
were {\em biased} random walkers whereas P ones were {\em pure} random
walkers.  Physically, $\alpha=0$ means that concentration gradient
and induced codeposition forces are exactly balanced by the electrical
force that goes in the opposite direction, towards the anode; besides,
P ion motion does not depend on Co ion motion.  With this
choice, we obtained qualitative agreement with the experimental results:
We reproduced the composition saturation and the homogeneity of the samples.
Nevertheless, the saturation
value of the dendritic alloys was wrong: The model predicted a value
around 92\% Co atoms.  We therefore gave up the (otherwise unrealistic)
hypothesis of exact
force balance and modify $\alpha$ to get quantitative agreement with the
experiments.  It turned out that $\alpha=0.3$ gave rise to the same
composition of the alloys as in the experiment, 88\% Co atoms.  It is
important to notice that this implies that P negative ions move towards
the negative electrode, and this can only come from the concentration
gradient and induced codeposition influences.

The simulation outcome (with $\alpha=0.3$) is summarized in Figs.\
\ref{Composition} to \ref{Homogeneous}.
Figure \ref{Composition} shows the dependence of the composition of the
alloys on the current density.  It can be seen from this plot that the
agreement between the model and the experiments is very good.  An
important conclusion that must be drawn from this plot is that there is
a direct relationship between the current density in the experiments and
the value of $p$.  It is to be expected that the exact coefficient will
depend on the type and the size of the lattice, but for the lattice we
are dealing with here, $J = p\times 5\times 10^6$ A m$^{-2}$.  After we
have determined the $p$ values that correspond to physical $J$ values
from the saturation curve, we are not free to modify $p$ anymore.
Hence, the $p$ values obtained from this relationship must also
reproduce the features of the aggregates grown with different $J$ if our
model is correct.  Besides, we have to fix the relationship between the
parameter $s$ and $p$. After trying several functional forms
\cite{newnota} we chose a simple linear dependence because it already
gave a very good agreement with the aggregate morphology.
This can be seen from Fig.\
\ref{Clusters}.  The aggregates in this figure were obtained in the
following form: A number of actual experimental values of $J$ and the
corresponding $T_{\scriptstyle\rm M}$ were taken.  From $J$, the values
of $p$ to be used in the simulations were computed through the above
relationship.  We then simulated the largest of them, $p=0.05$, for a
certain number of time steps.
This allowed us to obtain an equivalence between simulated
and actual time: We took the number of time steps in the simulation to
be proportional to the growth time for $J=2.5\times 10^5$ A m$^{-2}$
(i.e., $p=0.05$), which was 2 minutes.  By this procedure, we found
that 1 minute $= 1.4 \times 10^7$ simulation time steps.  Finally, we
simulated the rest of the cases stopping the simulation at a number of
time steps equivalent to the experimental, physical time.  The agreement
was fully satisfactory, and in all cases a number of particles of the
order of
12 000 was obtained.  Morphologies were also very similar to the ones
arising in the experiment, the ones grown with larger $p$ being more
open as shown in Fig.\ \ref{Clusters}.  Interestingly,
when the current
$J$ is very close to $0$, masses obtained in actual experiments (by
this and other groups) are rather small. This is so because the
deposit is not allowed to grow for a long enough time as would be
required by the relation $T_{\scriptstyle\rm M}\sim J^{-1}$. This is
shown in the top plot of Fig.\ \ref{Clusters}, an aggregate grown
for the equivalent of 60 minutes which would have needed twice that time
to reach $\sim$ 12 000 particles
($s$ was greater than it should be to induce fast growth).
In this small $J$ regime, the morphology of the aggregates becomes less
dendritic and more compact, reproducing the change reported in
\cite{MJ} from planar to dendritic shapes upon increasing the current
density.  Finally, Fig.\ \ref{Homogeneous} shows an
example of the cluster composition as a function of its height.  The
alloy bulk is fairly homogeneous, and fluctuations arise only at the
higher parts, where only the few particles at the growing tips
contribute.  The same degree of homogeneity was seen for all $J$
(equiv.\ $p$) values considered.

In summary, we have presented a simple model
that reproduces CoP growth by ECD in conditions very far from
equilibrium.  The main success of the model is that it can be compared
quantitatively to experiments, as the parameter $p$ and the simulation
time can be straightforwardly correlated to the physical $J$ and growth
time.  To our knowledge, this is the first time that such a quantitative
comparison is done.  The very good agreement in every aggregate
characteristic allows us to conclude that the model captures all the
essential physics of ECD in a very simple way.  It has to be noticed
then that in view of the model, the relevant processes must take place
in a thin layer in the neighborhood of the growing alloy, in agreement
with the considerations done in \cite{Fleury}.  As a final remark, we
want to point out that this model can be used to interpret other
experiments as well as to predict new effects that might arise when
changing the experimental conditions. Further experimental work is
needed to establish the validity limits of this model.

\bigskip

It is a pleasure to thank I.\ Cirac for valuable discussions.
A.S.\ is partially supported by DGICyT (Spain) through grant PB92-0378
and a MEC/Fulbright scholarship.  M.J.B.\ and J.M.R.\ acknowledge support
from DGICyT (Spain) grant MAT91-0031.  Work at Los Alamos is done under the
auspices of the U.S.\ D.o.E.

\begin{figure}
\caption[]{Alloy composition vs experimental current density $J$ and Co
bias $p$ in simulation. Full circles are obtained from simulation and
empty ones from experiment. Leftmost experimental and simulated points
lie on top of each other. Lines are only a guide to the
eye.}
\label{Composition}
\end{figure}

\begin{figure}
\caption[]{Clusters obtained for different currents and times.
Bottom to top: $J=2.5,\>
0.5,\> 0.1,\mbox{\rm\ and }
0.02 \times 10^5$ A m$^{-2}$ (equiv.\ $p=0.05,\>
0.01,\> 0.002,\mbox{\rm\ and } 0.0004$), whereas
$T_{\scriptstyle\rm M}=2,\> 10,\> 50$,\mbox{\rm\ and } 60 minutes (equiv.\
$T_{\scriptstyle\rm sim} = 2.8,\> 14,\> 70,\mbox{\rm\ and }
84\times 10^7$
time steps). Note that the morphology becomes more open when increasing
the current, becoming quite planar for the smallest $p$ value. The top
aggregate was grown with $s$ four times the one it should have according
to the linear relationship, to make it visible here. With the proper
$s$ value $T_{\scriptstyle\rm M}$ would be 300 minutes
(equiv.\ $350\times 10^7$ time steps).}
\label{Clusters}
\end{figure}

\begin{figure}
\caption[]{Alloy composition as a function of the aggregate coordinate
perpendicular to the cathode. Fluctuations at high values come from the
the fact that there are very few particles near the tips. This aggregate
was grown with $J=2\times 10^5$ A m$^{-2}$ ($p=0.04$) during 2.5 minutes
($3.5\times 10^7$ time steps).}
\label{Homogeneous}
\end{figure}


\begin{references}
\bibitem{otranota} New walkers were created above the top coordinate of
the growing aggregate. This constraint was removed in several
simulations to check whether three-dimensional effects could give rise
to different results. It was then possible that walkers were created
in closed empty regions of the aggregate. The outcome of the simulations
was very much similar to those carried out with the constraint.
\bibitem{nota} One can also think of choosing different $s$ for Co and
P ions. We did so and checked that the results changed very little even
taking rather different values.
\bibitem{newnota} Other dependences we tried were parabolic and hyperbolic
tangent, in different ranges of $p$. The obtained morphologies are
similar to the lineal one, so it seems enough to assume a smooth
increasing dependence.
\end{references}
\end{document}